\documentclass[twoside,amsmath,amssymb,fleqn]{article}
\usepackage[headings]{espcrc2}
\usepackage{natbib}
\usepackage{graphicx}

\newcommand{\var}{\mbox{Var}}
\newcommand{\xb}{\mbox{\bf x}}
\newcommand{\rb}{\mbox{\bf r}}
\newcommand{\ub}{\mbox{\bf u}}
\newcommand{\Ub}{\mbox{\bf U}}

\newcommand{\Rb}{\mbox{\bf R}}

\newcommand{\Db}{\mbox{\bf D}}
\newcommand{\Ab}{\mbox{\bf A}}
\newcommand{\Pc}{{\cal P}}
\newcommand{\del}{\mbox{\boldmath{$\nabla$}}}
\newcommand{\delp}{\del_\perp}
\newcommand{\Pe}{\mbox{P\hspace{-1pt}e\hspace{1pt}}}

\begin{document}


\title{ The Spot Model for random-packing dynamics }

\author{Martin Z. Bazant
\address{ 
Department of Mathematics, Massachusetts Institute of Technology,
Cambridge 02139 }
}

\runtitle{ The Spot Model for random-packings dynamics }
\runauthor{M. Z. Bazant}

%

\date{\today}

\begin{abstract}

The diffusion and flow of amorphous materials, such as glasses and
granular materials, has resisted a simple microscopic description,
analogous to defect theories for crystals.  Early models were based on
either gas-like inelastic collisions or crystal-like vacancy
diffusion, but here we propose a cooperative mechanism for dense
random-packing dynamics, based on diffusing ``spots'' of interstitial
free volume. Simulations with the Spot Model can efficiently generate
realistic flowing packings, and yet the model is simple enough for
mathematical analysis.  Starting from a non-local stochastic
differential equation, we derive continuum equations for tracer
diffusion, given the dynamics of free volume (spots). Throughout the
paper, we apply the model to granular drainage in a silo, and we also
briefly discuss glassy relaxation.  We conclude by discussing the
prospects of spot-based multiscale modeling and simulation of
amorphous materials.

\end{abstract}



\maketitle

\centerline{ Dedicated to Prager Medalist, Salvatore Torquato.}

\section{Introduction}

Professor Torquato has made pioneering contributions to the
characterization of random packings and their relation to properties
of heterogeneous materials~\citep{torquato}.  His recent work rejects
the classical notion of ``random close packing'' of hard spheres and
replaces it with the more precise concept of a ``maximally random
jammed state''~\citep{torquato00,torquato01,kansal02,donev04}. In
these studies and others~\citep{ohern02,ohern03}, however, the focus
is on the statistical geometry of {\it static} packings, and not on
the dynamics of nearly jammed packings in {\it flowing} amorphous
materials.

Dense random-packing dynamics is at the heart of condensed matter
physics, and yet it remains not fully understood at the microscopic
level. This is in contrast to dilute random systems (gases), where
Boltzmann's kinetic theory provides a successful statistical
description, based on the hypothesis of randomizing collisions for
individual particles. The same single-particle theory can also be
applied to molecular liquids at typical experimental time and length
scales, where kinetic energy is able to fully disrupt local
packings. The difficulty arises in describing liquids at very small
(atomic) length and time scales, over which the trajectories of
neighboring particles are strongly correlated -- the so-called {\it
cage effect} ~\citep{hansen}.

This difficulty is extended to much larger length and time scales in
glassy relaxation~\citep{angell00} and dense granular
flow~\citep{jaeger96}, where kinetic energy is insufficient to easily
tear a particle away from its cage of neighbors.  As a result, one
must somehow describe the cooperative motion of all particles at
once. In dense ordered materials (crystals), cooperative relaxation
and plastic flow are mediated by defects, such as interstitials,
vacancies, and dislocations, but it is not clear how to define
``defects'' for homogeneous disordered materials.

The challenge in describing random-packing dynamics is related to the
concept of ``hyper-uniform'' point distributions, recently introduced
by ~\citet{torquato03}. In a dilute gas, particles undergo independent
random walks and thus have the ``uniform'' distribution of a Poisson
process~\citep{hadji03}, where the variance of the number of
particles, $N$, scales with the volume, $V$: $\mbox{Var}(N) = \langle
N \rangle = \rho V$, where $\rho$ is the mean density. In a condensed
phase, the particle distribution must be ``hyper-uniform'', with much
smaller fluctuations, proportional to the surface area: $\mbox{Var}(N)
\propto V^{(d-1)/d}$, where $d=3$ is the spatial dimension, so it is
clear that particles cannot fluctuate independently.  In a crystal,
hyper-uniformity is a property of the ideal lattice, which is
preserved during diffusion and plastic flow by the motion of isolated
defects. For dense disordered materials, however, no simple flow
mechanism has been identified, which preserves hyper-uniformity.

\begin{figure}
\begin{center}
\includegraphics[width=1.5in]{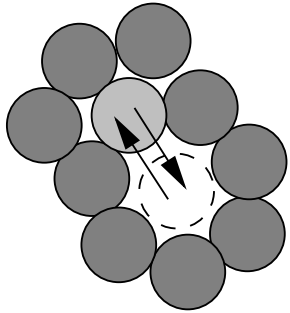} 
 \caption{
\label{fig:void}  Eyring's mechanism for flow in viscous liquids:  
A single particle jumps from one available ``cage'' to another by
exchanging with a ``void'' moving in the opposite direction. }
\end{center}
\end{figure}

\citet{eyring36} was perhaps the first to suggest a microscopic mechanism for
viscous flow in liquids, analogous to vacancy diffusion in
crystals. He proposed that the packing evolves when individual
particles jump into available cages, thus displacing pre-existing
voids, as shown in Fig.~\ref{fig:void}. Much later, the same
hypothesis was put forth independently in theories of the glass
transition~\citep{cohen59}, shear flow in metallic
glasses~\citep{spaepen77}, granular drainage from a
silo~\citep{lit58,mullins72}, and compaction in vibrated granular
materials~\citep{boutreux98}, although it is not clear that all of
these authors intended for the model to be taken literally at the
microscopic level. Since particles and voids simply switch places, it
seems the Void Model can only be simulated on a single, fixed
configuration of particles, but real flows are clearly not constrained
in this way.  This difficulty is apparent in the work of
\citet{hong91}, who neglected random packings and simulated
the Void Model on a lattice in an attempt to describe granular
drainage.

By now, free volume theories of amorphous materials (based on voids)
have fallen from favor, and experiments on glassy
relaxation~\citep{weeks00} and granular drainage~\citep{choi04} have
demonstrated that packing rearrangements are highly cooperative and
not due to single-particle hops. In a recent theory of granular chute
flows down inclined planes, \citet{ertas02} have postulated the
existence of coherent rotations, called ``granular eddies'' to
motivate a continuum theory of the mean flow. Although the theory
successfully predicts Bagnold rheology and the critical layer
thickness for flow, \citet{landry_note} have failed to find any
evidence for granular eddies in discrete-element simulations of chute
flow \citep{landry03}. 

In glass theory, \citet{adam65} introduced the concept of regions of
cooperative relaxation, whose sizes diverge at the glass
transition. Modern statistical mechanical approaches are based on
mode-coupling theory \citep{gotze92}, which accurately predicts
density correlation functions in simple liquids ~\citep{kob97}, although a
clear microscopic mechanism, which could be used in a particle
simulation, has not really emerged.
Cooperative rearrangements have also long been recognized in the
literature on sheared glasses. \citet{orowan52} was perhaps the first
to postulate localized shear transformations in regions of enhanced
atomic disorder.
\citet{argon79} later developed the idea of ``intense shear transformations'' at low
temperature, which underlies the stochastic model of ``localized
inelastic transformations''~\citep{bulatov94}. A similar notion of
``shear transformation zones'' (STZ) has also been invoked by
~\citet{falk98} in a continuum theory of shear response, which has
recently been extended to account for free-volume creation and
annihilation in glasses~\citep{lemaitre02} and granular
materials~\citep{lemaitre02c}. This phenomenology seems to capture
many universal features of amorphous materials, although the
microscopic picture of ``$+$'' and ``$-$'' STZ states remains vague.

In this paper, we propose a simple model for the kinematics of dense
random packings. In section~\ref{sec:spot}, we introduce a general
mechanism for structural rearrangements based on the concept of a
diffusing ``spot'' of free volume. In section~\ref{sec:gran}, we apply
the Spot Model to granular drainage. In section~\ref{sec:math}, we
analyze the diffusion of a tracer particle via a non-local, nonlinear
stochastic differential equation, in the limit of an ideal gas of
spots. In section~\ref{sec:graneq}, we derive equations for tracer
diffusion in granular drainage, which depend on the density, drift,
and diffusivity of spots (or free volume). We close by discussing
possible applications to glasses in section~\ref{sec:glass} and
spot-based multiscale modeling and simulation of amorphous materials
in section~\ref{sec:conc}.

\section{ The Spot Model }
\label{sec:spot}

\subsection{ Motivation }

Our intuition tells us that a particle in a dense random packing must
move together with its nearest neighbors over short distances,
followed by gradual cage breaking at longer distances. In simple
liquids, this transition occurs at the molecular scale ($<$ nm) over
very short times ($<$ ps) compared to typical experimental scales. In
supercooled liquids and glasses, the time scale for structural
relaxation effectively diverges and is replaced by slow, power-law
decay \citep{angell00,kob97,hansen}, although the length scale for
cooperative motion remains relatively small \citep{weeks00}. In
granular drainage, cage breaking occurs slowly, over time scales
comparable to the exit time from the silo, so that cooperative motion
is important throughout the system at the macroscopic scale
\citep{choi04}.

Another curious feature of granular drainage is the importance of
geometry: all fluctuations in a dense flow seem to have a universal
dependence on the distance dropped for a wide range of flow rates
\citep{choi04}. In a sense, therefore, increasing the flow speed in
this regime is like fast-forwarding a film, passing through the same
sequence of configurations, only more quickly. The only existing
theory to predict this property (as well as the mean flow profile in
silo drainage) is the Void Model, since increasing the flow speed
simply increases the injection rate of voids, but not their
geometrical trajectories. However, the model incorrectly predicts cage
breaking and mixing at the scale of individual particles. This
``paradox of granular diffusion'' is a fruitful starting
point for a new model of random-packing dynamics \citep{spotmodel}.

\subsection{ General formulation }

Let us suppose that the cage effect gives rise to spatial correlations
in particle velocities, with correlation coefficient, $C(r)$, for two
particles separated by $r$. More generally, when there is broken
symmetry, e.g. due to gravity in granular drainage, there is a
correlation coefficient, $C^{\alpha\beta}_p(\rb_1,\rb_2)$, for the
$\alpha$ velocity component of a particle at $\rb_1$ and the $\beta$
velocity component of a particle at $\rb_2$. Perhaps the simplest way
to encode this information in a microscopic model is to imagine that
particles move cooperatively in response to some extended entity --
{\it a spot} -- which causes a particle at $\rb_p$ to move by
\begin{equation}
\Delta\Rb_p = - w(\rb_p,\rb_s)\, \Delta\Rb_s  \label{eq:wdef}
\end{equation}
when it moves by $\Delta \Rb_s$ near $\rb_s$.  Although the spot is not
a ``defect'' {\it per se}, like a dislocation in an ordered packing,
it is a collective excitation which allows a random packing to
rearrange.

 
\begin{figure}
\begin{center}
\includegraphics[width=2in]{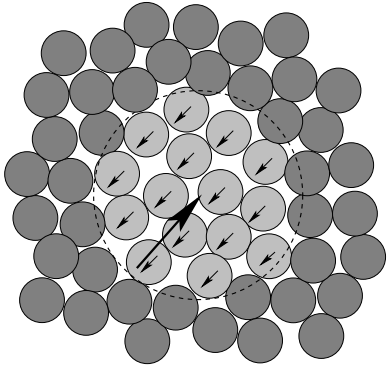}
\caption{ The spot 
mechanism for cooperative diffusion: A group of neighboring particles
makes small correlated displacements in response to a diffusing
``spot'' of excess interstitial volume. \label{fig:spot}
}
\end{center}
\end{figure}

In principle, the spot influence, $w$, could be a matrix causing a
smooth distribution of local translation and rotation about the spot
center $\rb_s$, but a reasonable first approximation is that it is
simply a collective translation the opposite direction from the spot
displacement, as illustrated in Fig.~\ref{fig:spot}. In this case, $w$
is a scalar function, whose shape is roughly that of the velocity
correlation function. More precisely, under some simple assumptions,
we show below that $C(r)$ is the overlap integral of two spot
influence functions separated by $r$. Since the spot influence is
related to the cage effect, we expect that $w$ and $C$ will decay
quickly with distance, for separations larger than a few particle
diameters. Due to the local statistical regularity of dense random
packings, we might expect a spot to retain its ``shape'' as it moves,
in which case $w$ depend only on the separation vector, $\rb_p^{(i)} -
\rb_s^{(j)}$, although this assumption might need to be relaxed in
regions of large gradients in density or velocity.

Physically, what is a spot? Since particles move collectively in one
direction, a spot must correspond to some amount of {\it free
interstitial volume} (or ``missing particles'') moving in the other
direction. If particles are distributed with number density
$\rho_p(\rb_p)$ and a spot at $\rb_s$ carries a typical volume,
$V_s(\rb_s)$, then an approximate statement of volume conservation is
\begin{equation}
V_s \Delta \Rb_s = - \int d\rb_p  \rho(\rb_p)
w(\rb_p,\rb_s) \Delta \Rb_p(\rb_p). 
\label{eq:volcons}
\end{equation}
For particles distributed uniformly at volume fraction, $\phi$, this
reduces to a simple expression for the local volume carried by a spot,
\begin{equation}
V_s(\rb_s) = \phi \int d\rb_p w(\rb_p,\rb_s),  \label{eq:svol}
\end{equation}
which can thus be interpreted as a measure of the spot's ``total
influence''.

 
\begin{figure}
\begin{center}
\includegraphics[width=2.5in]{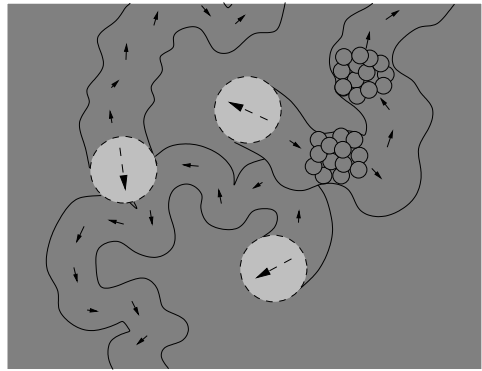}
\caption{ The trails of spots
correspond to transient, reptating chains of
particles. \label{fig:trails} }
\end{center}
\end{figure}

A very simple, spot-based Monte Carlo simulation proceeds as
follows. Given a distribution of (passive) particles and (active)
spots, the random displacement, $\Delta \Rb_s^{(j)}$, of the $j$th
spot centered at $\rb_s^{(j)}$ induces a random displacement,
$\Delta\Rb_p^{(i)}$, of the $i$th particle centered at $\rb_p^{(i)}$:
\begin{equation}
\Delta \Rb_p^{(i)} = - w(\rb_p^{(i)},\rb_s^{(j)}+\Delta\Rb_s^{(j)}) \,
\Delta\Rb_s^{(j)}  \label{eq:discrete}
\end{equation}
Each spot undergoes an independent random walk, with an appropriate
drift and diffusivity for free volume, which leaves in its trail a
thick chain of particles reptating in the opposite direction, as shown
in Fig.~\ref{fig:trails}. In Eq.~(\ref{eq:discrete}), we choose to
center the spot influence on the end of its small displacement, but it
is also reasonable to use the midpoint of the
displacement\citep{ssim}. In the infinitesimal limit (see below),
these choices are analogous to different definitions of stochastic
differentials \citep{risken}.

In principle, the drift velocity, diffusivity, and influence function
of spots could depend on local variables, such as stress, and
temperature (or rather, some suitable microscopic quantities related
to contact forces and velocities, respectively). Spots could also interact
with each other, undergo creation and annihilation, and possess a
statistical distribution of sizes (or influence functions).  The
simplest kinematic assumption, however, which captures the basic
physics of the cage effect, is that spots are identical and maintain
their properties while undergoing independent (non-interacting) random
walks. In particular, the constant influence function,
$w(|\rb_p-\rb_s|)$, is chosen to be translationally invariant in space
and time.  It turns out that this model allows rather realistic multiscale
simulations, while remaining analytically tractable.

\begin{figure*}
\begin{center}
\includegraphics[width=5in]{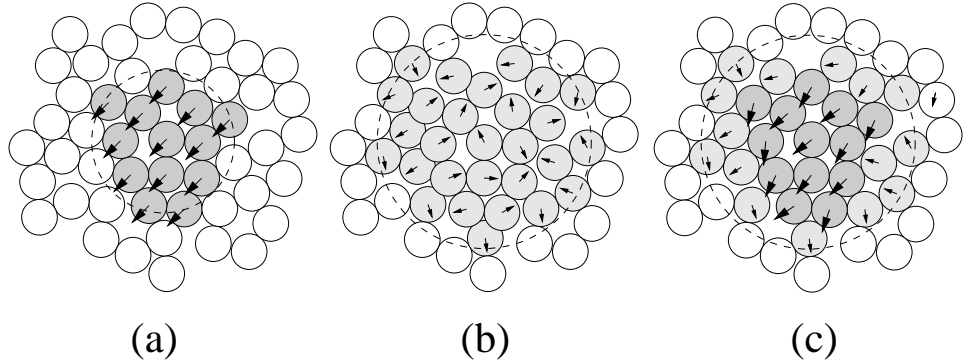}
\caption{ Multiscale simulation of densely packed (nearly)
hard spheres with the Spot Model: (a) A block of neighboring particles
translates opposite to the displacement of a spot of free volume; (b)
the block and a shell of neighbors are allowed to relax under
soft-core repulsive forces; (c) the net cooperative rearrangement
combines these two steps. (Particle displacements are greatly
exaggerated for clarity.)  \label{fig:relax} }
\end{center}
\end{figure*}

\subsection{ Multiscale simulation  }

The simple spot mechanism above gives a reasonable description of
tracer diffusion and slow cage breaking in random packings, but it
does not strictly enforce packing constraints (or, more generally,
inter-particle forces). As such, particles perform independent random
walks in the long-time limit, which eventually leads to uniform
density fluctuations with Poisson statistics. For a complete
microscopic model, we must somehow preserve hyper-uniform
packings.

This may be accomplished by adding a relaxation step to the
spot-induced displacements, as shown in Fig.~\ref{fig:relax}.  First in (a),
a spot displacement causes a simple correlated displacement, as
described above, e.g. using Eq.~(\ref{eq:discrete}) with some
short-ranged choice of $w(r)$ with a finite cutoff. Next in (b), the
affected particles and a shell of their nearest neighbors are allowed
to relax under appropriate inter-particle forces, with more distant
particles held fixed. For simulations of (nearly) hard grains, the
most important forces come from a soft-core repulsion, which pushes
particles apart only if particle begin to overlap. 

Although it is not obvious {\it a priori}, the net spot-induced
cooperative displacements, shown in
Fig.~\ref{fig:relax}(c), easily produce very realistic flowing
packings, while preserving the physical picture of the model
\citep{ssim}. In practice, the correlated nature and small size of the
spot-induced block displacements results in very small and infrequent
particle overlaps, only near the edges of the spot, where some shear
occurs with the background packing. As a result, it seems the details
of the relaxation are not very important, although this issue merits
further investigation. In any case, the algorithm is
interesting in its own right as a method of {\it multiscale
modeling}, since it combines a macroscopic simulation of simple
extended objects (spots) with localized, microscopic simulations of
particles. 

\section{ Application to Granular Drainage}
\label{sec:gran}

\subsection{ Spot parameters}

The classical Kinematic Model for the mean velocity in granular
drainage \citep{nedderman}, which compares fairly well with
experiments \citep{tuzun79,samadani99,choiexpt}, postulates
that the mean downward velocity, $v$, satisfies a linear diffusion
equation,
\begin{equation}
\frac{\partial v}{\partial z} = b \, \delp^2 v \label{eq:kin}
\end{equation}
where the vertical distance $z$ plays the role of ``time'' and the
horizontal dimensions (with gradient, $\delp$) play the role of
``space''. The microscopic justification for Eq.~(\ref{eq:kin}) is the
Void Model of \citet{lit58,lit63b} and \citet{mullins72,mullins74},
where particle-sized voids perform directed random walks upward from
the orifice. As discussed above, this microscopic mechanism is firmly
contradicted by particle-tracking experiments \citep{choi04}, but, as
shown below, a similar macroscopic flow equation can be derived from
the Spot Model, where spot diffuse upward with a (horizontal)
diffusion length,
\begin{equation}
b = \frac{\var(\Delta\xb_s)}{2d_h\Delta z_s}  \label{eq:bspot}
\end{equation}
where $\Delta\xb_s$ is the random horizontal displacement of a spot as
it rises by $\Delta z_s$ and $d_h=2$ is the horizontal dimension. A
typical value for 3 mm glass beads is $b \approx 1.3 d$, where $d$ is
the particle diameter.

The shape of the spot influence function can be inferred from
measurements of spatial velocity correlations in experiments
\citep{spotmodel} or simulations by the discrete-element method
(DEM) \citep{ssim}. The simplest assumption is a uniform spherical
influence with a finite cutoff,
\begin{equation}
w(r) = \left\{ \begin{array}{ll}
w & r< d_s/2 \\
0 & r > d_s/2
\end{array} \right.
\end{equation}
where experiments and simulations find $d_s \approx 5 d$. This is
consistent with our interpretation of the spot mechanism in terms of
the cage effect, where a particle moves with its nearest
neighbors. The typical number of particles affected by a spot,
\begin{equation}
N =  \phi \left(\frac{d_s}{d}\right)^3,
\end{equation}
is thus $N \approx 72$, for $\phi
\approx 0.58$.

For a uniform spot, the condition of volume conservation,
Eq.~(\ref{eq:volcons}), reads
\begin{equation}
V_s\, (\Delta \xb_s, \Delta z_p) = - N \, V_p \, (\Delta \xb_p,\Delta
z_p) . \label{eq:spotsimple} 
\end{equation}
Using Eq.~(\ref{eq:wdef}), this provides an expression for the spot influence,
\begin{equation}
w = \frac{V_s}{N V_p} \approx \frac{\Delta\phi}{\phi^2}  \label{eq:wsimple}
\end{equation}
in terms of $\Delta\phi$, the change in local volume fraction due to
the presence of a single spot.  In discrete-element simulations of
granular spheres in silo drainage \citep{ssim}, the local volume
fraction varies in the range, $\phi = 0.565-0.605$, within the rough
bounds of jamming, $\phi =0.63$
~\citep{torquato00,kansal02,ohern02,ohern03}, and random loose
packing, $\phi=0.55$ ~\citep{onoda90}. If we attribute
$\Delta\phi/\phi=1\%$ to a single spot, then we find $w\approx
0.017$, but, if many spots, say $N_s=10$, can overlap, then this
estimate is reduced by $1/N_s$. We thus expect, $w = 10^{-3} -
10^{-2}$, which can be tested against experiments.

We can also use Eqs.~(\ref{eq:spotsimple})-(\ref{eq:wsimple}) to infer
$w$ from a measurement of the horizontal particle diffusion length: 
\begin{equation}
b_p = \frac{\var(\Delta\xb_p)}{2d_h|\Delta z_p|} =
 \frac{w^2 \var(\Delta\xb_s)}{2d_h w \Delta z_s} = w\, b  \label{eq:bpsimple}
\end{equation}
Discrete-element simulations \citep{ssim} and particle-tracking
experiments \citep{choi04} for similar flows yield $w = b_p/b =
0.00286 d/1.14 d = 0.00250$ and $w = d/b\Pe_x = d/(1.3 d)(321) =
0.0024$, respectively (where $\Pe_x$ is a P\'eclet number). These
values are consistent with the predictions of the model.

A spot's influence is related to the free volume it carries by
Eq.~(\ref{eq:svol}). In the case of a uniform spot of diameter, $d_s$,
its total free volume is given by
\begin{equation}
V_s = \frac{\pi \phi w d_s^3}{6} 
\end{equation}
which is related to the particle volume, $V_p$, by
\begin{equation}
\frac{V_s}{V_p} = w \phi \left(\frac{d_s}{d}\right)^3 = w N.
\end{equation}
For dense granular drainage, the typical values $N=72$ and $w=0.0025$
imply $V_s = 0.18 V_p$, so a spot carries only around one fifth of a
particle volume, spread out over a region of roughly five particle
diameters.  This delocalized picture of free volume diffusion is
radically different from the classical Void Model, which is the
(unphysical) limit where $V_s=V_p$ and $N=w=1$.

\begin{figure*}
\begin{center}
(a)\includegraphics[width=2.4in]{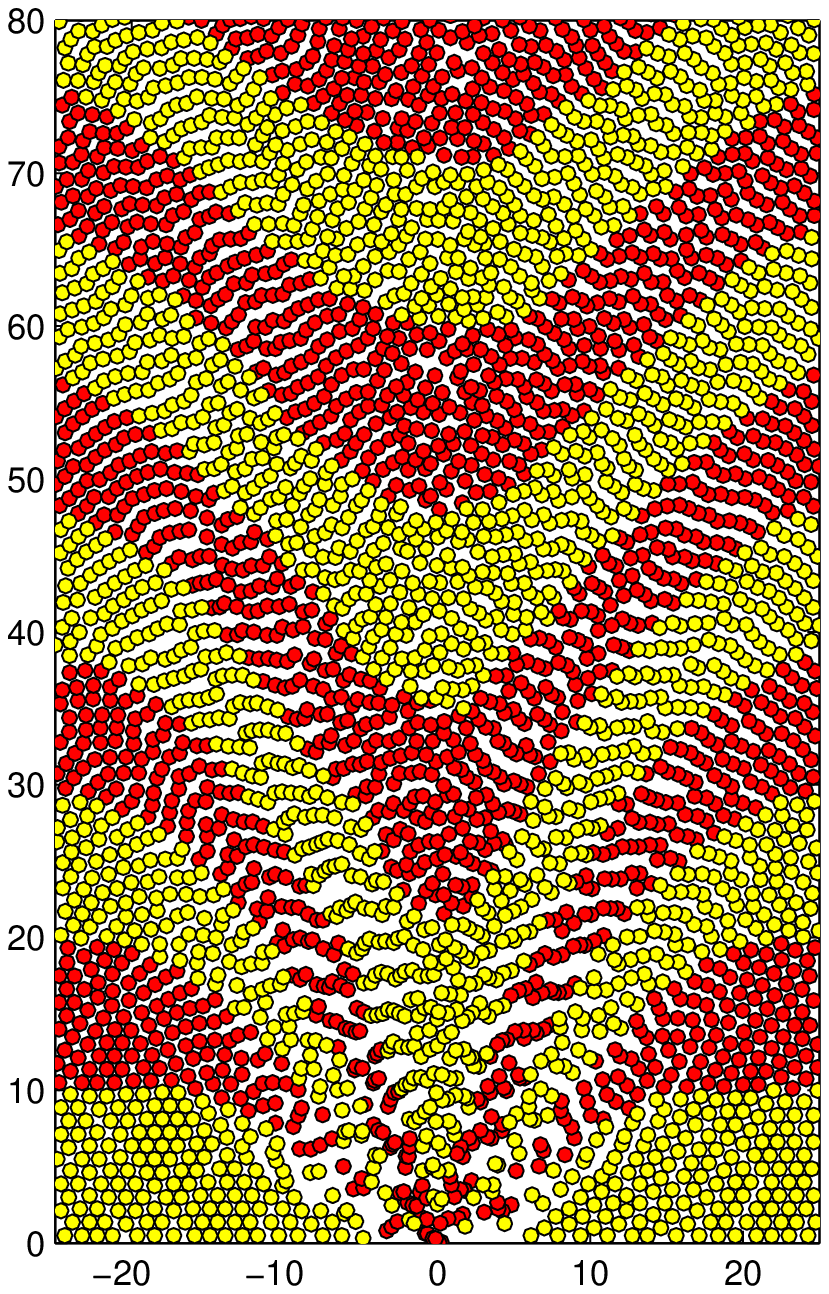} \nolinebreak
(b)\includegraphics[width=2.4in]{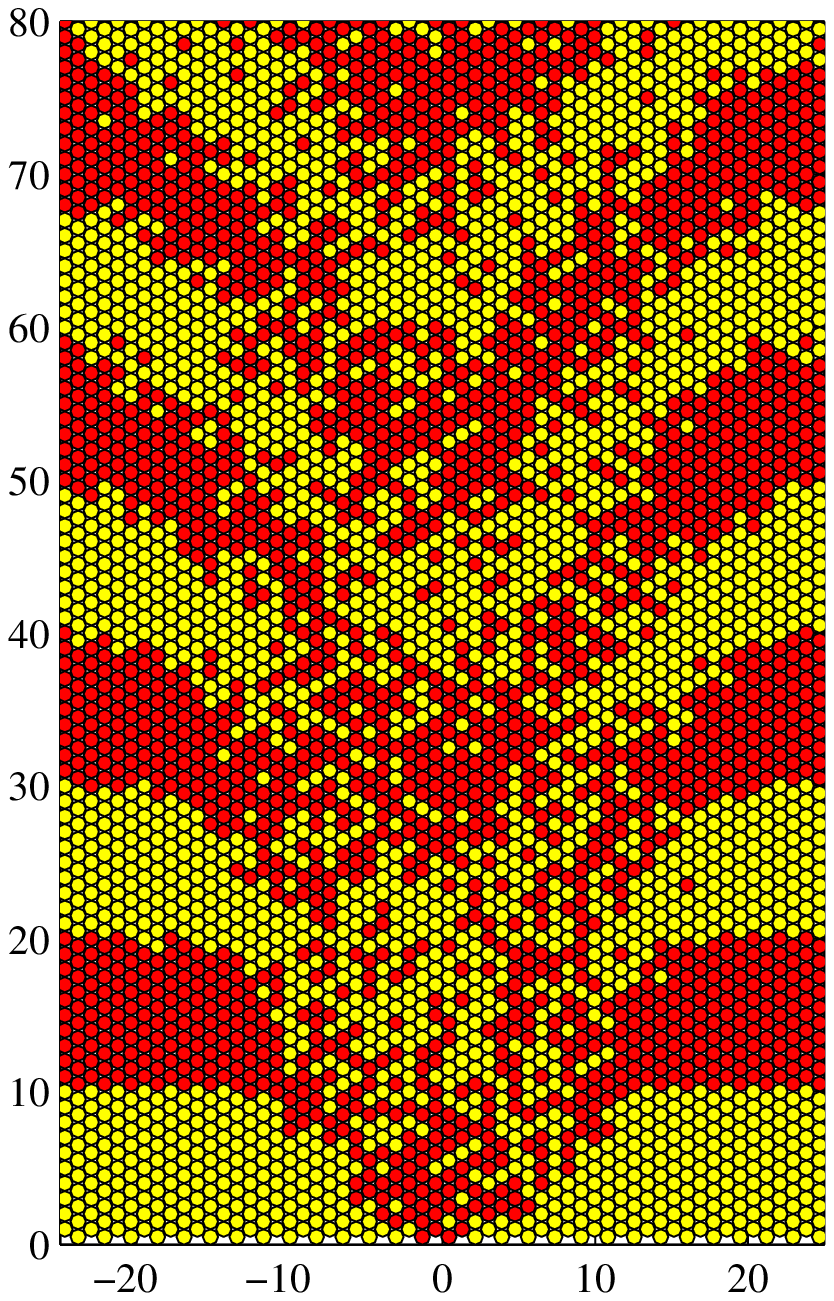} \\
(c)\includegraphics[width=2.4in]{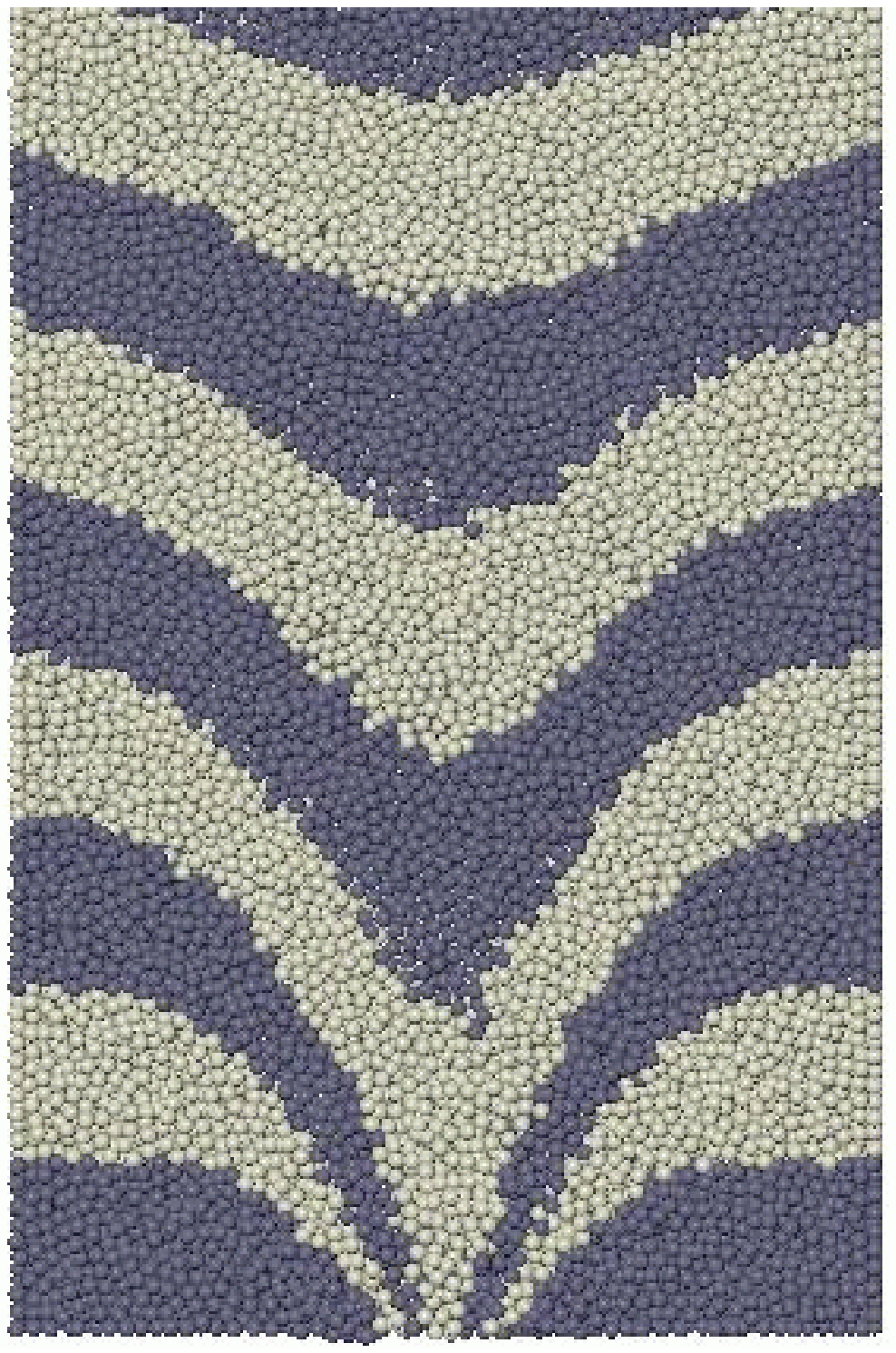} \nolinebreak
(d)\includegraphics[width=2.4in]{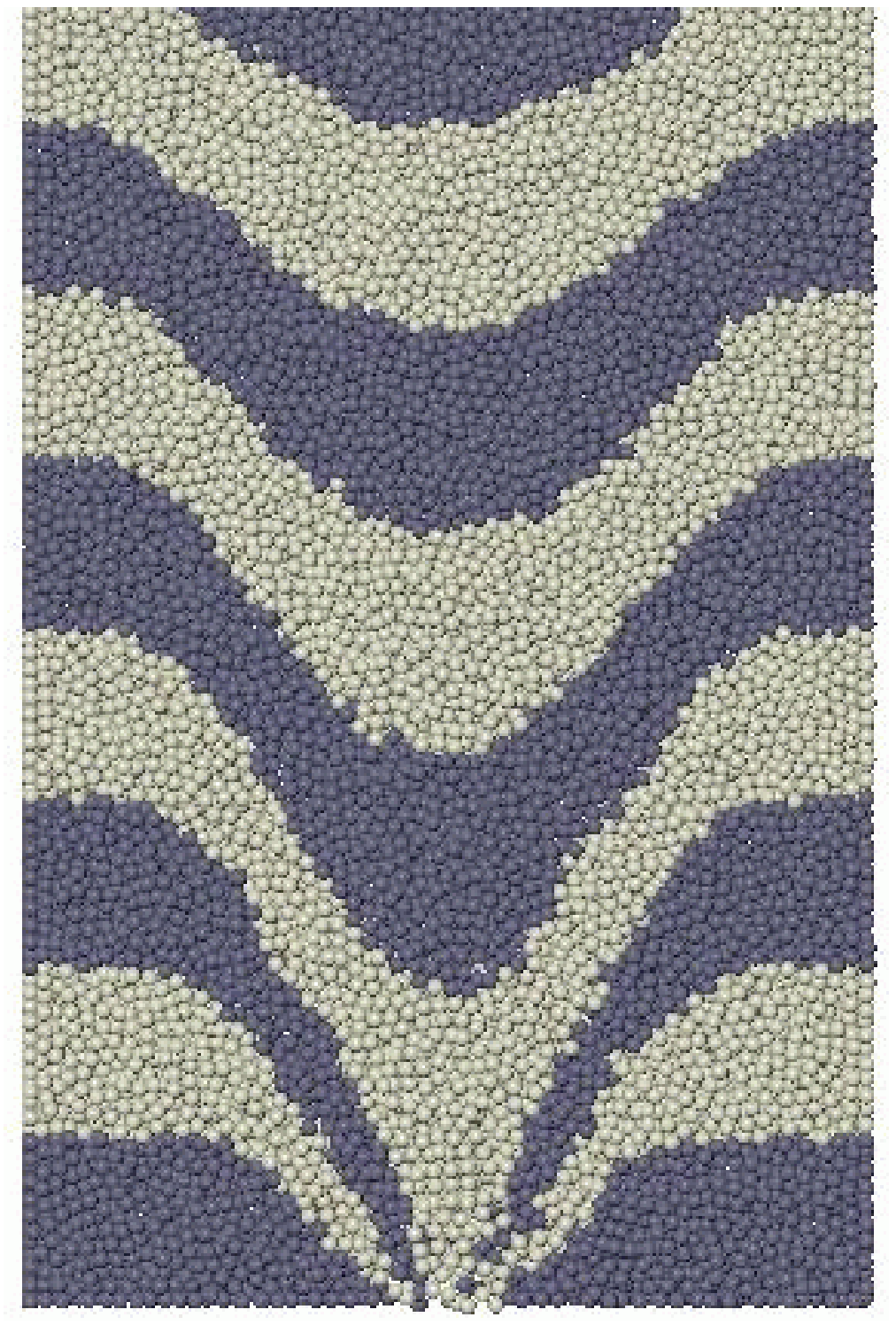}
\caption{ Simulations of granular drainage in a quasi-two-dimensional
silo. Top: Two-dimensional simulations using (a) the Spot Model
without packing constraints and (b) the Void Model
\citep{spotmodel}. Bottom: Three-dimensional simulations using (c) the
Spot Model with multiscale relaxation and (d) the Discrete Element
Method for frictional, visco-elastic spheres \citep{ssim}. Particles
are colored according to their initial positions in horizontal stripes, $10 d$ thick.
\label{fig:sim} }
\end{center}
\end{figure*}

\subsection{ Spot simulations }

As described above, the Spot Model can be successfully calibrated for
granular drainage to enable very simple and efficient
simulations. First, we consider the basic mechanism in
Fig.~\ref{fig:spot} and Eq.~(\ref{eq:discrete}) applied to granular
drainage in a quasi-two-dimensional silo with a narrow opening, far
from the side walls, as in the experiments of \citet{samadani99} and
\citet{choi04}. For simplicity, we simulate the model in two
dimensions ($d_h=1$) using uniform spots with the parameters
determined from experiments above ($w=0.0024$, $d_s=5d$, $b=1.3
d$). The simulation begins with a random packing of identical disks,
colored with horizontal stripes ($10 d$ thick) to aid in visualizing
the subsequent evolution. 

A snapshot of the spot simulation at a later time is shown in
Fig.~\ref{fig:sim}(a). For comparison, a simulation of the same
situation with the Void Model on a two-dimensional lattice, following 
\citet{hong91}, is shown in Fig.~\ref{fig:sim}(b), along with the central slice
of a three-dimensional DEM simulation ($15 d$ thick) in
Fig.~\ref{fig:sim}(d), which is very similar to the
experiment. Although the mean flow profile is similar in the spot and
void simulations and reasonably close to experiment, the void
simulation displays far too much diffusion and cage breaking, since
the initial horizontal stripes are completely mixed down to the single
particle level inside the flow region. In contrast, the interfaces
between the colored layers remain fairly sharp in the spot simulation,
as in experiments and the DEM simulations. Tracer diffusion is
described fairly well by the spot simulation, although the lack of
packing constraints eventually leads to the loss of hyper-uniformity,
as particles begin to overlap and open gaps in the lower side regions
of highest shear. For more details, see \citet{spotmodel}.

Next, we consider the multiscale spot algorithm in
Fig.~\ref{fig:relax}, with an internal relaxation step,
applied to a three-dimensional  drainage simulation, starting from the same
initial condition and geometry as the DEM simulation in
Fig.~\ref{fig:sim}(d).  The simplest possible relaxation scheme is to
push any pair of overlapping particles in a relaxation zone (a sphere
of diameter, $d_s + 4d$) apart by a displacement, $\alpha(d-r)$
proportional to the overlap, $d-r$, while keeping particles fixed
outside a sphere of diameter, $d_s+2d$. A spot
simulation with similar parameters as above, including such a
relaxation step with $\alpha=0.8$, is shown in
Fig.~\ref{fig:sim}(c). The rate of introducing spots at the orifice
and their upward drift velocity has also been calibrated for
comparison to the DEM simulation, at the same instant in time.

Clearly, the simple multiscale relaxation step is able to preserve
realistic random packings, and in many ways the spot
simulation in Fig.~\ref{fig:sim}(c) is indistinguishable from the
much more computationally demanding simulation in
Fig.~\ref{fig:sim}(d).  Not only are the mean velocity profile and
diffusion length reproduced, but so are various microscopic statistics
of the packing geometry, such as the two-body and three-body
correlation functions \citep{ssim}.  These surprising results are quite insensitive
to the details of the relaxation step, apparently due to the very
small and infrequent particle overlaps which arise from the
cooperative mechanism in the spot simulation. Another, deeper reason
may be that the geometry of dense flowing random packings has
universal features, which are achieved by the totally different
dynamics of spot and DEM simulations. 

\section{ Mathematical analysis of diffusion }
\label{sec:math}

\subsection{ A stochastic differential equation }

In this section, we return to the general formulation of the Spot
Model and analyze tracer diffusion in the continuum limit. It is clear
from the simulations in Fig.~\ref{fig:sim} that the basic model in
Fig.~\ref{fig:spot} gives a reasonable description of the dynamics of
a single particle tracer, even though the multiscale relaxation step
in Fig.~\ref{fig:relax} is needed to preserve realistic packings.  The
relatively small size of the relaxation displacements makes it
reasonable to regard them a small additional ``noise'' in a
mathematical analysis of tracer diffusion.  Here, we will neglect this
small (but complicated) noise and view its average effect as
incorporated statistically into the spot influence function,
$w(\rb_p,\rb_s)$, in Eq.~(\ref{eq:discrete}).

\begin{figure}
\begin{center}
\includegraphics[width=2.2in]{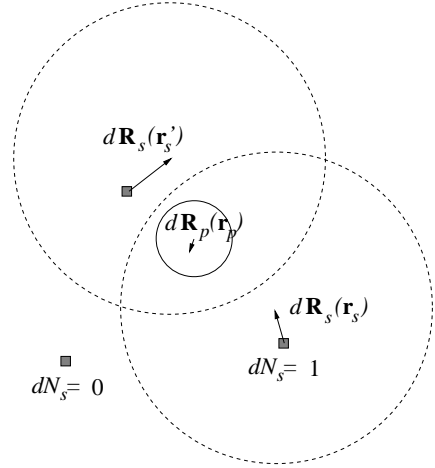}
\caption{ \label{fig:SDE_defs} Sketch of a particle interacting
with a collection of passing spots, showing some of the quantities
involved in the nonlocal SDE, Eq. ~ (\protect\ref{eq:tracer}). }
\end{center}
\end{figure}

We begin by partitioning space as shown Fig.~\ref{fig:SDE_defs}, where
the $n$th volume element, $\Delta V_s^{(n)}$, centered at
$\rb_s^{(n)}$ contains a random number, $\Delta N_s^{(n)}$, of spots
at time $t$ (typically one or zero). In a time interval, $\Delta t$,
suppose that the $j$th spot in the $n$th volume element makes a random
displacement, $\Delta \Rb_s^{(j,n)}$ (which could be zero).
According to Eq.~(\ref{eq:discrete}), the total displacement, $\Delta
\Rb_p$, of a particle at $\rb_p$ in time $\Delta t$ is then given by a
sum of all the random displacements induced by nearby moving spots,
\begin{eqnarray} 
 \Delta\Rb_p =  - \sum_n 
\sum_{j=1}^{\Delta N_s^{(n)}} 
   w(\rb_p, \rb_s^{(n)} +
\Delta\Rb_s^{(j,n)}) 
\Delta\Rb_s^{(j,n)}.  \nonumber \\
  \label{eq:tracer_discrete} 
\end{eqnarray} 
Note that the spatio-temporal distribution of spots, $\Delta
N_s^{(n)}$, is another source of randomness, in addition to the
individual spot displacements, $\Delta\Rb_s^{(j,n)}$, so that each
particle displacement is given by a {\it random sum of random variables}.

In the limit of infinitesimal displacements, we arrive at a
non-local, nonlinear stochastic differential equation (SDE):
\begin{eqnarray}
d\Rb_p(t) &=&  - \int dN_s(\rb_s,t)   \label{eq:tracer} \\
& &  w(\rb_p(t), \rb_s + d\Rb_s(\rb_s,t)) \nonumber
\, d\Rb_s(\rb_s,t), 
\end{eqnarray}
where the stochastic integral is defined by the usual limit
(infinitely refined partition of space) of the random Riemann sum in
Eq.~(\ref{eq:tracer_discrete}).  This equation differs from standard
nonlinear SDEs \citep{risken} in two basic ways: (i) The tracer
trajectory,
\begin{equation}
\rb_p(t) = \int_{\tau=0}^t d\Rb_p(\tau)   \label{eq:trajectory}
\end{equation}
is passively driven by a stochastic distribution of moving influences
(spots), $dN_s(\rb_s,t)$, which evolves in time and space, rather than
by some internal source of independent noise, and (ii)
the stochastic differential, $d\Rb_p(t)$, is given by a
non-local integral over other stochastic differentials, $d\Rb_s(\rb_s,t)$,
associated with these moving influences, which lie at positions,
$\rb_s$, at finite distances away from the particle at $\rb_p$.

\subsection{ A Fokker-Planck equation  }

In general, the various stochastic differentials in
Equation~(\ref{eq:tracer}) are correlated, which significantly
complicates analysis. Here, we will make the reasonable first
approximation of an {\it ideal gas of spots}, where the tracer
particle sees an independent, random configuration of non-interacting
spots at each infinitesimal time step. As in an ideal gas
\citep{hadji03}, spots are thus distributed according to a
Poisson process with a given mean density, $\rho_s(\rb_s,t)$. In
addition to neglecting correlations caused by interactions between
spots, we disregard the following facts: (i) The distribution of spots
in space, $\{dN_s(\rb_s,t)\}$, at time $t$ depends explicitly on the
distribution and displacements at the previous time, $t - dt$, via the
spot random walks; (ii) Each spot, due to its finite range of
influence, affects the same particle for a finite period of time, so
any persistence (autocorrelation) in the spot trajectory is
transferred to the particles, in a nonlinear fashion controlled by
$w(\rb_p,\rb_s)$.

In the spot-gas approximation, the tracer particle performs a random
walk with independent (but non-identically distributed) displacements,
which depend non-locally on a Poisson process for finding
spots. Therefore, the propagator, $P_p(\rb,t|\rb_0,t_0)$, which gives
the probability density of finding the particle at $\rb$ at time $t$
after being at $\rb_0$ at time $t_0$, satisfies the following
Fokker-Planck equation~\citep{risken},
\begin{equation}
\frac{\partial P_p}{\partial t} + \del\cdot(\ub_p P_p)
= \del\del : (\Db_p P_p)   \label{eq:fp} ,
\end{equation}
with drift velocity,
\begin{equation}
\ub_p(\rb,t) = \frac{\langle d\Rb_p(\rb,t)\rangle}{dt} = \lim_{\Delta
  t \rightarrow 0} \frac{\langle \Delta \Rb_p(\rb,t) \rangle}{\Delta
  t} ,
\end{equation}
and  diffusivity tensor, 
\begin{equation}
\Db_{p}^{\alpha\beta}(\rb,t) = \frac{\langle dR_p^{\alpha} 
dR_p^{\beta} \rangle}{2\,  dt} .
\end{equation}
(Here $\del\del : \Ab$ denotes $\sum_\alpha \sum_\beta
\frac{\partial^2 A^{\alpha\beta}}{\partial {x_\alpha} \partial
{x_\beta}}$.)  The Fokker-Planck coefficients may be calculated by
taking the appropriate expectations using
Eq.~(\ref{eq:tracer_discrete}) in the limits $\Delta V_s^{(n)}
\rightarrow 0$ and $\Delta t \rightarrow 0$ (in that order), which is
straightforward since we assume that spots do not interact. Here, the
spot displacements, $\Delta\Rb_s^{(j)}(\rb_s^{(n)})$, and the local
numbers of spots, $\Delta N_s^{(n)}$, are independent random variables
in each time interval, and they are independent of the same
variables at earlier times.

In order to calculate the drift velocity, we need only the mean spot
density, $\rho_s(\rb_s,t)$, defined by $\langle \Delta N_s^{(n)}
\rangle = \rho_s(\rb_s,t) \, \Delta V_s^{(n)}$.  The result,
\begin{eqnarray}
\ub_p(\rb_p,t) &=& - \int dV_s \, w(\rb_p,\rb_s) \,
\left[ \rho_s(\rb_s,t) \ub_s(\rb_s,t) \right. \nonumber \\
& &  \left. - 2\, \Db_s(\rb_s,t) \cdot \del \rho_s(\rb_s,t) \right] \label{eq:up}
\end{eqnarray}
exhibits two sources of drift. The first term in the integrand is a
particle drift velocity, which opposes the spot drift velocity,
\begin{equation}
\ub_s(\rb,t) = \frac{\langle d\Rb_s(\rb,t)\rangle}{dt}.
\end{equation}
as in Eq.~(\ref{eq:spotsimple}). 
The second term, which depends on the spot diffusion tensor,
\begin{equation}
\Db_{s}^{(i,j)}(\rb,t) = \frac{\langle dR_s^{(i)} dR_s^{(j)}\rangle}{2\,
    dt} ,
\end{equation}
is a ``noise-induced drift'', typical of nonlinear
SDEs~\citep{risken}, which causes particles to climb gradients in the
spot density. This extra drift is crucial to ensure that particles
eventually move toward the source of spots, e.g. the orifice in
granular drainage. Both contributions to the drift velocity in
Eq.~(\ref{eq:up}) are averaged non-locally over a finite region,
weighted by the spot influence function, $w(\rb_p,\rb_s)$.

In order to calculate the diffusivity tensor, we also need information
about fluctuations in the spot density. From the spot-gas
approximation, we have
\[
\langle \Delta N_s^{(n)} \Delta N_s^{(m)} \rangle =
\delta_{m,n} \langle (\Delta N_s^{(n)})^2\rangle = O((\Delta
V_s^{(n)})^\nu), 
\]
where $\nu=1$ for a Poisson process and
$\nu<1$ for a hyper-uniform process~\citep{torquato03}.  It turns out 
that such fluctuations do not contribute to the diffusion
tensor (in more than one dimension), and the result is 
\begin{eqnarray}
\Db_p(\rb_p,t) = \int dV_s  w(\rb_p,\rb_s)^2 \rho_s(\rb_s,t) 
\Db_s(\rb_s,t). \nonumber \\
  \label{eq:Dp}
\end{eqnarray}
Note that the influence function, $w$, appears squared in
Eq.~(\ref{eq:Dp}) and linearly in Eq.~(\ref{eq:up}), which causes the
P\'eclet number for tracer particles to be of order $w$ smaller than
that of spots (or free volume), as in Eq.~(\ref{eq:bpsimple}).

Higher-order terms a Kramers-Moyall expansion generalizing
Eq.~(\ref{eq:fp}) for finite independent displacements, which do depend
on fluctuations in the spot density, are straightforward to calculate,
but beyond the scope of this paper.  Such terms are usually ignored
because, in spite of improving the approximation, they tend to produce
small negative probabilities in the tails of
distributions~\citep{risken}. In granular materials, however, velocity
gradients can be highly localized, so the correction terms could be
useful.

\subsection{ Spatial velocity correlation tensor }

For any stochastic process representing the motion of a single
particle, it is well-known that transport coefficients can be
expressed in terms of {\it temporal} correlation functions via the
Green-Kubo relations~\citep{risken}. For example, the diffusivity
tensor in a uniform flow is given by the time integral of the velocity
auto-correlation tensor,
\begin{equation}
D_p^{\alpha\beta} = \int_0^\infty dt \langle U_p^\alpha(t) U_p^\beta(0)
\rangle 
\end{equation}
where $\Ub_p(t) = \{ U_p^\alpha\} = d\Rb_p/dt$ is the stochastic
velocity of a particle. (A similar relation holds for spots.) 

In the Spot Model, nearby particles move cooperatively, so the
transport properties of the collective system also depend on the
two-point {\it spatial} velocity correlation tensor,
\begin{equation}
C_p^{\alpha\beta}(\rb_1,\rb_2) =
\frac{\langle U_p^\alpha(\rb_1) U_p^\beta(\rb_2) \rangle} 
{\sqrt{\langle U_p^\alpha(\rb_1)^2\rangle \langle
U_p^\beta(\rb_2)^2 \rangle }}   \label{eq:Cdef}
\end{equation}
which is normalized so that $C_p^{\alpha\beta}(\rb,\rb)=1$.  We
emphasize that the expectation above is conditional on finding two
particles at $\rb_1$ and $\rb_2$ at a given moment in time and
includes averaging over all possible spot distributions and
displacements.  Substituting the SDE (\ref{eq:tracer}) into
Eq.~(\ref{eq:Cdef}) yields
\begin{eqnarray}
C_p^{\alpha\beta}(\rb_1,\rb_2) &= \int dV_s \, \rho_s(\rb_s)    
w(\rb_1,\rb_s)  w(\rb_2,\rb_s)  \nonumber \\
&  D_s^{\alpha\beta}(\rb_s) /  \sqrt{ D_p^{\alpha\beta}(\rb_1)  D_p^{\alpha\beta}(\rb_2)}  \label{eq:Cp}
\end{eqnarray} 
assuming independent spot displacements.

Equation~(\ref{eq:Cp}) is an integral relation for
cooperative diffusion, which relates the spatial velocity correlation
tensor to the spot (or free volume) diffusivity tensor via integrals
of the spot influence function, $w(\rb_p,\rb_s)$.  If the statistical
dynamics of spots is homogeneous (in particular, if $\Db_s$ is
constant), then the relation simplifies:
\begin{eqnarray}
& C^{\alpha\beta}_p(\rb_1,\rb_2) =   \\
& \frac{\int dV_s \, \rho_s(\rb_s) \,
w(\rb_1,\rb_s) \, w(\rb_2,\rb_s) }
{\sqrt{ \int dV_s \, \rho_s(\rb_s) \, w(\rb_1,\rb_s)^2 \,
 \int dV_s^\prime \, \rho_s(\rb_s^\prime) \, w(\rb_2,\rb_s^\prime)^2 
}}. \nonumber
\end{eqnarray}
if also the tensor is diagonal, $D_s^{\alpha\beta} \propto
\delta_{\alpha,\beta}$.  If the statistical dynamics of particles is
also homogeneous, as in a uniform flow ($\rho_s=$ constant), then it
simplifies even further:
\begin{equation}
C^{\alpha\beta}_p(\rb) = \frac{\int dV_s \, w(\rb-\rb_s) 
\, w(-\rb_s) } { \int dV_s \, w(\rb_s)^2 }
\end{equation}
where we have assumed that the spot influence function, and thus the
correlation tensor, is translationally invariant ($\rb =
\rb_1-\rb_2$).  In this limit, as mentioned above, the velocity
correlation function is simply given by the (normalized) overlap integral for
spot influences separated by $\rb$.


\subsection{ Relative diffusion of two tracers }

The spatial velocity correlation function affects many-body transport
properties. For example, the relative displacement of two tracer
particles, $\rb = \rb_1 - \rb_2$, has an associated diffusivity
tensor given by,
\begin{eqnarray}
D^{\alpha\beta}(\rb_1,\rb_2)& =& D_p^{\alpha\beta}(\rb_1) +
D_p^{\alpha\beta}(\rb_2) \\
& &  - 2 \, C^{\alpha\beta}_p(\rb_1,\rb_2)\,
\sqrt{ D_p^{\alpha\alpha}(\rb_1)  D_p^{\beta\beta}(\rb_2) }  \nonumber
\end{eqnarray}
In  a uniform flow, the diagonal components take the simple form
\begin{equation}
D^{\alpha\alpha}(\rb) = 2\, D^{\alpha\alpha}_p \left( 1 -
C^{\alpha\alpha}_p(\rb)\right) 
\end{equation}
which may be used above to estimate the cage-breaking time, as the
expected time for two particles diffuse apart by more than one
particle diameter. A more detailed calculation of the relative
propagator, $P(\rb,t|\rb_0,t_0)$, neglecting temporal correlations (as
above) would start from the associated Fokker-Planck equation,
\begin{equation}
\frac{\partial P(\rb,t)}{\partial t} = \del\del : \left(\Db(\rb)
P(\rb,t)\right)  \label{eq:fprel}
\end{equation}
with a delta-function initial condition. (In a non-uniform flow, one
must also account for noise-induced drift and motion of the the center
of mass.) This analysis does not enforce packing constraints, so it
allows for two particles to be separated by less than one diameter. A
hard-sphere repulsion may be approximated by a reflecting boundary
condition at $|\rb|=d$ when solving equations such as
(\ref{eq:fprel}), but there does not seem to be any simple way to
enforce inter-particle forces exactly in the analysis.

\section{ Tracer diffusion in granular drainage }
\label{sec:graneq}

\subsection{ Statistical dynamics of spots }

The analysis in the previous section makes no assumptions about spots,
other than the existence of well-defined local mean density, mean
velocity, and diffusion tensor, which may depend on time and space.
As such, the results may have relevance for a variety of dense
disordered systems exhibiting cooperative diffusion (see below). In
this section, we apply the model to the specific case of granular
drainage, in which spots diffuse upward from a silo orifice, as in
Fig.~\ref{fig:sim}. Our goal here is simply to show how to derive
continuum equations from the Spot Model in a particular case, but not
to study any solutions in detail. 

For simplicity, let us assume that each spot undergoes mathematical Brownian
motion with a vertical drift velocity, $\ub_s = v_s \hat{z}$,
and a diagonal diffusion tensor,
\begin{equation}
\Db_s = \left( \begin{array}{ccc}
D_s^\perp & 0 & 0 \\
0 & D_s^\perp  & 0 \\
0 & 0 & D_s^\|
\end{array} \right)   \label{eq:Ds}
\end{equation}
which allows for a different diffusivity in the horizontal ($\perp$)
and vertical ($\|$) directions due to  symmetry breaking 
by  gravity. In that case, the propagator for a single
``spot tracer'',
$P_s(\xb,z,t|\xb_0,z_0,t)$, satisfies another Fokker-Planck equation,
\begin{eqnarray}
\frac{\partial P_s}{\partial t} + \frac{\partial }{\partial
z}\left(v_s P_s\right) =
\del^2_\perp \left( D_s^\perp P_s \right) +
 \frac{\partial^2 }{\partial z^2}\left(D_s^\| P_s\right) . \nonumber \\
 \label{eq:spotfp} 
\end{eqnarray} 
The coefficients may depend on space (e.g. larger velocity above the
orifice than near the stagnant region), as suggested by the shape of
some experimental density waves ~\citep{behringer89}.

The geometrical spot propagator, $\Pc_s(\xb,|z,\xb_0,z_0)$, is the
conditional probability of finding a spot at horizontal position $\xb$
once it has risen to a height $z$ from an initial position
$(\xb_0,z_0)$. For constant $v_s$ and $\Db_s$, the geometrical
propagator satisfies the diffusion equation, 
\begin{equation}
\frac{\partial \Pc_s}{\partial z} = b\, \del^2_\perp \Pc_s \label{eq:spotdiff}
\end{equation} 
where $b = D_s^\perp v_s$ is the kinematic parameter.  If spots move
independently, this equation is also satisfied by the steady-state
mean spot density, $\rho_s(\xb,z)$, analogous to Eq.~(\ref{eq:kin}) of
the Kinematic Model. However, the mean particle
velocity in the Spot Model, Eq.~(\ref{eq:up}), is somewhat different,
as it involves nonlocal effects (see below).

The time-dependent mean density of spots, $\rho_s(\xb,z,t)$, depends
on the mean spot injection rate, $Q(\xb_0,z_0,t)$
(number/area$\times$time), which may vary in time and space due to
complicated effects such as arching and jamming near the orifice. It
is natural to assume that spots are injected at random points along
the orifice (where they fit) according to a space-time Poisson process
with mean rate, $Q$. In that case, if spots do not interact, the
spatial distribution of spots within the silo at time $t$ is also a
Poisson process with mean density,
\begin{eqnarray}
\rho_s(\xb,z,t) &=& \int d\xb_0 \int dz_0 \int_{t_0<t} dt_0  \label{eq:rhos} \\
& & Q(\xb_0,z_0,t)\, P_s(\xb,z,t|\xb_0,z_0,t_0) .  \nonumber
\end{eqnarray}
For a point-source of spots (i.e. an orifice roughly one spot wide)
at the origin with flow rate, $Q_0(t)$ (number/time), this reduces to
\begin{eqnarray}
\rho_s(\xb,z,t) = \int_{t_0<t} dt_0 \, Q_0(t_0)\, P_s(\xb,z,t|0,0,t_0)
, \nonumber \\
\ 
\end{eqnarray}
where $P_s$ is the usual Gaussian propagator for Eq.~(\ref{eq:spotfp})
in the case of constant $u_s$ and $\Db_s$. In reality, spots should
weakly interact, but the success of the Kinematic Model suggests that
spots diffuse independently as a first approximation in granular
drainage \citep{choiexpt}.

\subsection{ Statistical dynamics of particles }

Integral formulae for the drift velocity and diffusivity tensor of a
tracer particle may be obtained by substituting the spot density which
solves Eq.~(\ref{eq:rhos}) into the general expressions ~(\ref{eq:up})
and (\ref{eq:Dp}), respectively.  For example, if spots only diffuse
horizontally ($D_s^\|=0$), then the mean downward velocity of
particles is given by
\begin{equation}
v_p(\rb,t) =  \int dV_s \, w(\rb_p,\rb_s)\, \rho_s(\rb_s,t) \, v_s(\rb_s,t)
\end{equation}
Note that the mean particle velocity is a nonlocal average of nearby spot
drift velocities. 

For simplicity, let us consider a bulk region where the
spot density varies on scales much larger than the spot size. In this
limit, the integrals over the spot influence function reduce to the
following ``interaction volumes'':
\begin{equation}
V_k(\rb) = \int d\rb_s w(\rb,\rb_s)^k
\end{equation}
for $k=1,2$. (Note that $V_1=V_s$ above.) The equation
for tracer-particle dynamics (\ref{eq:fp}) then takes the form,
\begin{eqnarray} 
\frac{\partial P_p}{\partial t} &=& \frac{\partial }{\partial
z}\left[\left( v_s \rho_s - 2 D_s^\| \frac{\partial \rho_s}{\partial z}
\right) V_1 P_p \right]  \label{eq:spotdrainfp} 
 \\
& &  - 2 \del_\perp\cdot\left( D_s^\perp (\del_\perp \rho_s)
V_1 P_p\right)  
\nonumber \\
& & 
 \frac{\partial^2 }{\partial z^2}\left(D_s^\| \rho_s V_2 P_p \right)
+
\del^2_\perp \left( D_s^\perp \rho_s V_2 P_p \right). \nonumber 
\end{eqnarray} 
Again, it is clear that rescaling the spot density is equivalent to
rescaling time.

When the spot dynamics is homogeneous (i.e. $u_s$ and $\Db_s$ are
constants), Equation~(\ref{eq:spotdrainfp}) simplifies further:
\begin{eqnarray}
\frac{1}{v_s V_s} 
\frac{\partial P_p}{\partial t} &=& \left( \frac{\partial }{\partial
z} + b_p^{\perp} \del^2 + b_p^{\|}
 \frac{\partial^2 }{\partial z^2}\right)(\rho_s P_p) \ \ \  \label{eq:simplespotfp} 
\\
 & &  - 2 b^\perp \del\cdot(P_p \del \rho_s) 
- 2 b^\| \frac{\partial}{\partial z}\left(P_p \frac{\partial
\rho_s}{\partial z}\right)  \nonumber 
\end{eqnarray} 
where $b^\perp = b = D_s^\perp/v_s$ and $b^\| = D_s^\|/v_s$ are the
spot diffusion lengths and $b_p^\perp = b_p V_2/V_1$ and $b_p^{\|} =
b^{\|} V_2/V_1$ are the particle diffusion lengths. In this
approximation, the latter are given by the simple formula,
\begin{equation}
\frac{b_p^{\perp}}{b^\perp} = \frac{b_p^\|}{b^\|} 
= \frac{\int dV_s \, w(\rb,\rb_s)^2 }{\int dV_s w(\rb,\rb_s)}  \label{eq:bp}
\end{equation}
which generalizes Eq.~(\ref{eq:bpsimple}) for a uniform spot with a
sharp cutoff. 
The physical meaning of the diffusion lengths becomes more clear in
the limit of uniform flow, $\rho_s =$ constant. In terms of the
position in a frame moving with the mean flow, $\zeta = v_p t - z$,
where $v_p = v_s V_s \rho_s$, we arrive at a simple diffusion
equation,
\begin{equation}
\frac{\partial P_p}{\partial \zeta} = \left( b_p^{\perp} \del_\perp^2 +
b_p^{\|} \frac{\partial^2}{\partial z^2}\right) P_p ,
\end{equation}
where $\zeta$, the mean distance dropped, acts like time, consistent
with the experimental findings of \citet{choi04}.

\section{ Possible application to glasses }
 \label{sec:glass}

We have seen that the Spot Model, in its simplest form, accurately
reproduces the kinematics of bulk granular drainage, so it is tempting
to speculate that it might be extended to flows in other amorphous
materials. In this section, we briefly consider evidence for spot-like
dynamics in glasses, but we leave further extensions of the Spot Model
for future work.

Experiments have revealed ample signs of ``dynamical heterogeneity''
in supercooled liquids and glasses~\citep{hansen,kob97,angell00}, but
the direct observation of cooperative motion has been achieved only
recently.  Rather than compact regions of relaxation, \citet{glotzer98}
have observed ``string-like'' relaxation in molecular dynamics
simulations of a Lennard-Jones model glass. The strength and length
scale of correlations increases with decreasing
temperature, consistent with the Adam-Gibbs
hypothesis. 
Such cooperative motion would be difficult to observe experimentally
in a molecular glass, but \citet{weeks00} have used confocal
microscopy to reveal three-dimensional clusters of faster-moving
particles in a dense colloids. In the supercooled liquid phase,
clusters of cooperative relaxation have widely varying sizes, which
grow as the glass transition is approached. In the glass phase, the
clusters are much smaller, on the order of ten particles, and do not
produce significant rearrangements on experimental time scales.

These observations suggest that the Spot Model may have relevance for
structural rearrangements in simple glasses.  String-like relaxation
is reminiscent of the trail of a spot, in Fig.~\ref{fig:trails}. An
atomically thin chain might result from the random walk of a spot,
roughly one particle in size, but carrying less than one particle of
free volume. Larger regions of correlated motion might involve larger
spots and/or collections of interacting spots. Some key features of
the experimental data of~\citet{weeks00} seem to support this idea:
(i) Correlations take the form of ``neighboring particles moving in
parallel directions'', as in Fig.~\ref{fig:spot}; and (ii) the large
clusters of correlated motion tend to be fractals of dimension two,
as would be expected for the random-walk trail of a spot, as in
Fig.~\ref{fig:spot}(b). For a complete theory of the glass transition,
however, one would presumably have to consider interactions between
spots and thermal activation of their creation, motion, and
annihilation.

\section{ Conclusion }
\label{sec:conc}

In this paper, we have introduced a mechanism for structural
rearrangements of dense random packings, due to diffusing spots of
free volume. Even without inter-particle forces, the Spot Model gives
a reasonable description of tracer dynamics, which is trivial to
simulate and amenable to mathematical analysis, starting from a
nonlocal stochastic differential equation. With a simple relaxation
step to enforce packing constraints, the Spot Model can efficiently
produce very realistic flowing packings, as demonstrated by the case
of granular drainage from a silo. The spot mechanism may also have
relevance for glassy relaxation and other phenomena in amorphous
materials.

Regardless of various material-specific applications, the ability to
easily produce three-dimensional dense random packings is interesting
in and of itself. Current state-of-the-art algorithms to generate
dense random packings are artificial and computationally 
expensive, especially near jamming
\citep{torquato00,kansal02,ohern02,ohern03}. A popular example is the
molecular dynamics algorithm of
\cite{lubachevsky90}, which simulates a dilute system of interacting
particles, whose size grows linearly in time until jamming occurs. For
each random packing generated, however, a separate molecular dynamics
simulation must be performed. In contrast, the Spot Model produces a
multitude of dense random packings (albeit with some correlations
between samples) from a single simulation, which is more efficient
than molecular dynamics, since it does not require
the mechanical relaxation of all particles at once.  It would be interesting to
characterize the types of dense packings generated by the Spot Model
and  compare with the results of other algorithms.

The Spot Model (with relaxation) also provides a convenient paradigm
for multiscale modeling and simulation of amorphous materials,
analogous to defect-based modeling of crystals. The iteration between
global ``mesoscopic'' simulation of spots and local ``microscopic''
simulation of particles leads to a tremendous savings in computational
effort, as long as the spot dynamics is physically realistic for a
given system. For example, a simple extension of the multiscale
simulations of granular spheres by \citet{ssim} would be to different
particle shapes, such as ellipsoids, which have been shown to pack
more efficiently than spheres~\citep{donev04b}. The only change in the
simulation would be to modify the inter-particle forces in the
relaxation step for a soft-core repulsion with a different shape.

A more challenging and fruitful extension would be to incorporate
mechanics into the multiscale simulation, beyond geometrical packing
constraints. One way to do this may be use the information about
inter-particle forces in the spot relaxation step to estimate local
stresses, which could then affect the dynamics of spots. It may also be
necessary to move particles directly in response to mechanical forces,
in addition to the random cooperative displacements caused by
spots. Such extensions seem necessary to describe forced shear flows
in granular and glassy materials. For now, at least we have a
reasonable model for the kinematics of random packings.

\vspace{0.2in}

\section*{Acknowledgements}

This work was supported by the U. S. Department of Energy (grant
DE-FG02-02ER25530) and the Norbert Wiener Research Fund and NEC Fund
at MIT.  The author is grateful to J. Choi, A. Kudrolli,
R. R. Rosales, C. H. Rycroft for many stimulating discussions and to
A. S. Argon, L. Bocquet, M. Demkowicz, R. Raghavan for references to
the glass literature.

\bibliography{spotmodel}

\end{document}